\documentstyle[12pt]{article}

\def\a{\begin{eqnarray}}
\def\b{\end{eqnarray}}
\def\0{\nonumber}
\def\l{\lambda}

\def\s{\sigma}
\def\al{\alpha}

\def\ep{\epsilon}
\def\om{\omega}
\def\co{\Omega}
\def\th{\theta}
\def\var{\varphi}

\input amssym.def
\font\teneusm=eusm10                    
\font\seveneusm=eusm7                   
\font\fiveeusm=eusm5                    
\newfam\eusmfam
\textfont\eusmfam=\teneusm
\scriptfont\eusmfam=\seveneusm
\scriptscriptfont\eusmfam=\fiveeusm

\def\sG{sine--Gordon}
\def\shG{sinh--Gordon}
\def\dg{ dressing group}

\def\aTo { affine Toda }
\def\aTe {affine Toda equations}

\def\CaT {Conformal affine Toda}
\def\dg{ dressing group}

\def\dges{ dressing group elements}

\def\dt{ dressing transformation}
\def\dts{ dressing transformations}

\def\lg {loop group}

\renewcommand{\theequation}{\thesection.\arabic{equation}}

\newlength{\extraspace}
\newlength{\extraspaces}
\newcounter{dummy}

\newcommand{\ai}{
\addtocounter{equation}{1}
\setcounter{dummy}{\value{equation}}
\setcounter{equation}{0}
\renewcommand{\theequation}{\thesection.\arabic{dummy}\alph{equation}}
\begin{eqnarray}
\addtolength{\abovedisplayskip}{\extraspaces}
\addtolength{\belowdisplayskip}{\extraspaces}
\addtolength{\abovedisplayshortskip}{\extraspace}
\addtolength{\belowdisplayshortskip}{\extraspace}}
\newcommand{\bj}{
\end{eqnarray}
\setcounter{equation}{\value{dummy}}
\renewcommand{\theequation}{\thesection.\arabic{equation}}}

\input amssym
\newcommand{\bq}{\begin{equation}}
\newcommand{\eq}{\end{equation}}
\newcommand{\ba}{\begin{eqnarray}}
\newcommand{\ea}{\end{eqnarray}}
\newcommand{\ban}{\begin{eqnarray*}}
\newcommand{\ean}{\end{eqnarray*}}
\newcommand{\brr}{\begin{array}}
\newcommand{\err}{\end{array}}
\newcommand{\bc}{\begin{center}}
\newcommand{\ec}{\end{center}}

\def\A{{\cal A}}

\def\E{{\cal E}}
\def\H{{\cal H}}
\def\G{{\cal G}}

\def\S{{\cal S}}

\def\P{{\cal P}}

\def\T{{\cal T}}

\def\l{\lambda}

\def\s{\sigma}
\def\al{\alpha}

\def\ep{\epsilon}
\def\o{\omega}
\def\co{\Omega}
\def\th{\theta}
\def\var{\varphi}
\def\An{{A_n^{(1)}}}
\def\sl{{sl(n+1)}}
\def\sll{{\widetilde{sl}(n+1)}}
\def\Sll{{\widetilde{SL}(n+1)}}

\newcommand{\bea}{\begin{eqnarray}}
\newcommand{\eea}{\end{eqnarray}}
\newcommand{\bean}{\begin{eqnarray*}}
\newcommand{\eean}{\end{eqnarray*}}
\newcommand{\CC}{\Bbb C}
\newcommand{\PP}{\Bbb P}
\newcommand{\ZZ}{\Bbb Z}

\newcommand{\del}{\partial}

\begin{document}

\begin{titlepage}

\begin{flushright}
CBPF.NF.059/96 \\
hep--th/9612029
 \end{flushright}

\vskip0.5cm
\centerline{\LARGE   $A_n^{(1)}$ Toda solitons and the dressing symmetry }
\vskip1.5cm
\centerline{\large   H. Belich
\footnote{E--mail address belich@cbpfsu1.cat.cbpf.br}     
 and R. Paunov
\footnote{E--mail address paunov@cbpfsu1.cat.cbpf.br}}
\centerline{Centro Brasileiro de Pesquisas Fisicas }
\centerline{Rua Dr. Xavier Sigaud 150, Rio de Janeiro, Brazil}

\vskip5cm
\abstract{ We present an elementary derivation of the soliton--like 
solutions in the $A_n^{(1)}$  Toda models which is alternative to the
previously used Hirota method. 
The solutions of the underlying linear problem   
corresponding to the $N$--solitons are calculated. This enables 
us to obtain explicit expression for the element
which by dressing group action, produces a generic soliton 
solution. In the particular example of monosolitons we 
suggest a relation to the vertex operator formalism, 
previously used by Olive, Turok and Underwood. Our results 
can also   be considered as generalization of the approach 
to the sine--Gordon solitons, proposed by Babelon and Bernard. }

\end{titlepage}

\section{ Introduction}

\setcounter{equation}{0}
\setcounter{footnote}{0}

The solitons appear in various topics of the modern mathematical 
and elementary particle physics \cite{Raj}. Generically, the 
solitons represent  field configurations which are topologically
 non--trivial. Therefore they turn  to be a natural object in
 studying nonperturbative effects in  the quantized
theory \cite{Raj}, \cite{FaKo}. There is an another (non--equivalent) 
approach 
to the solitons which treats them as solutions of integrable 
non--linear evolution equations \cite{Ab}, \cite{Nov}, \cite{Fad}. 
The crucial point is 
that the equations of motion appear as a zero
 curvature condition of
 a certain 
Lax connection. As a consequence of this, one can map,
 in a manner explained in
\cite{Ab}--\cite{Fad}, the original dynamical 
variables into a set of 
scattering data,
which due to the integrability, satisfy linear
 evolution equations.
This transformation, called direct spectral 
transformation,
yields the action--angle variables
 of the system.  To get back
the original dynamical variables one has 
to perform the so--called
 inverse scattering transformation. 
This approach, known in 
the literature as the
Inverse Scattering Method (ISM) \cite{Ab}--\cite{Fad} provides 
an elegant procedure
to solve integrable non--linear 
evolution equations.

Within the ISM, the solitons arise after imposing the 
vanishing of the reflection coefficient of
 the corresponding
 Lax operator. Due to the last condition, 
the inverse scattering equations 
reduce to a linear algebraic system, which only
 depends on the scattering data
related to the discrete spectrum 
of the Lax operator. 

The Toda equations \cite{T} admit a Lax representation 
\cite{OT}, and due to the existence of a classical 
$r$--matrix, are integrable. In \cite{OT} another
 important idea has  been advanced:
 to use generalized Cartan matrices
in order to obtain integrable field equations.
 In particular, the \aTo\, theories correspond 
to the extended Cartan matrices of simple Lie algebras. 
The last are derived from the 
simple Lie algebra Cartan matrices by adding 
the extended root which is  minus the highest  root. 
There is an alternative approach \cite{DS} which
 consists in the study of the properties of a Lax
 operator which has  a special form. The ideas
 developed in \cite{DS} have been further explored in 
\cite{gDS} to get generalized Drinfeld--Sokolov 
integrable hierarchies. The crucial step within this
 approach is to expand, after a suitable gauge
 transformation, the components of the Lax connection
 into certain maximally 
commuting subalgebra of the underlying affine 
Lie algebra.
 This ensures the existence of infinite number 
of conserved quantities.  Similar procedure 
applied to the \aTo\, models appeared also in
 \cite{eng}.

The \dg\, is a special symmetry of the integrable 
evolution equations 
\cite{ZS}. It admits an elegant interpretation
 within the tau--function approach \cite{jap}.
 Dressing transformations act on the components 
of the corresponding
Lax connection without changing its form \cite{fr}. 
Therefore, the \dg\, turns out to be a symmetry 
of the phase space. To ensure the covariance of the 
symplectic structure under dressing transformations, 
one has  to introduce a specific Poisson bracket on 
the \dg\,  \cite{fr}, \cite{STS}. In this paper we 
are not going to analyze this problem since it 
 presents difficulties  even for the \dges\, which
 generate monosolitons from the vacuum
in the \sG\, 
theory \cite{GR}. The dressing symmetry has  been 
exploited recently
\cite{ago} to treat a huge class of integrable
 hierarchies which admit a vacuum
solution.  
          
The soliton solutions in the $\An$  Toda model 
 are found by Hollowood \cite{Hol} who used the 
Hirota method. Later it became clear that the 
Hirota method can be applied to get solitons   
in arbitrary \aTo\, models 
\cite{Clis}. 
The relation between the soliton solutions in
 the \aTo\, theories
and special elements in the affine (or Kac--Moody) 
Lie algebras, called vertex operators is clarified 
in \cite{Olive}. An intriguing property of the
  formalism, developed in these papers is that
 it permits     generalization for the Toda
 theories based on twisted affine Lie algebras 
\cite{Cat}.

The actual interest to the  affine Toda solitons 
is due to their relation
to the $N$--body integrable systems \cite{body}.
 It turns out that the solitons are related to
 certain (relativistically invariant) integrable
 systems with finite degrees of freedom. Another
 interesting property of the affine Toda solitons
 is that they are closely related to the
 Seiberg--Witten duality \cite{dual}.

We outline the content of this paper. In Sec. 2 we
 generalize a method to get
solitons introduced by Date \cite{Date} which is
 complementary both to the 
Hirota method \cite{Hol}, \cite{Clis} and to the ISM
\cite{Nid}.
To illustrate the generality of our approach we fix
 the coupling constant to be real. This wants to say
 that we are working with algebraic instead of the
 usually treated physical solitons which only exist
for imaginary values of the coupling constant.
As a 
result we obtain the $N$--solitons as they were
 expressed in \cite{body}. In Sec. 3 we evaluate 
the \dg\, element which generates solitons 
from the vacuum in the defining representation of
 $sl(n+1)$. Our result appears to be a generalization 
of the element calculated by Babelon and Bernard \cite{BB}. 
Sec. 4 is devoted to the relation to the vertex operator
 construction of solitons advanced by Olive, Turok and 
Underwood \cite{Olive}.

\section{ Soliton solutions in the $\An$ Toda theories.}

\setcounter{equation}{0}
\setcounter{footnote}{0}

The problem of finding soliton solutions in the affine Toda 
models was previously approached in the literature by using 
the Hirota method \cite{Hol}, \cite{Clis} and with the help 
of group theoretical methods \cite{Olive}. 
Soliton solutions only exist for 
imaginary values of the coupling constant. Moreover, it was 
clarified that despite the equations of motion and the 
lagrangian density are complex, the 
solitons carry {\it real} momentum and energy. 
The properties of 
the \aTo\, systems, both for real and imaginary 
coupling constants are studied 
within the ISM \cite{Nid}. The  standard 
scheme of the ISM meets certain obstructions when 
applied to the \aTo\,models based on arbitrary 
simple Lie algebras. In particular, it turns out that 
the Jost solutions \cite{Fad}, and therefore the
 elements of the transition matrix,  loose their nice 
analiticity properties as functions on the
 spectral parameter. In this Section we generalize
 an elegant method \cite{Date} to obtain the 
$\An$ Toda solitons. It exploites two important 
features of the soliton 
solutions: first, due to the vanishing of the 
reflection coefficient of the 
auxiliary linear problem, the corresponding Jost 
solutions are single valued 
functions on the spectral parameter $\l$; second, 
the soliton Jost solutions
are uniquely determined by the scattering data 
related to the discrete spectrum
of the underlying Lax operator. Applying the ideas 
developed in \cite{Date}
to the $\An$ Toda models, we  recover the equations 
describing the discrete spectrum of the 
corresponding linear problem, by using a finite order 
inner automorphism $\s$ of the simple Lie 
algebra $A_n$. The last has order $n+1$ and
 introduces the so called {\it principal gradation}
 in the affine Lie 
algebra $\An$ \cite{Kac}.

The $\An$ Toda equations are equivalent to a 
zero--curvature condition of 
a connection, the components of which belong to
 the loop Lie algebra
$\sll$ in the principal gradation. We will start by 
introducing certain basic facts concerning the 
Lie algebras $sl(n+1)$, 
$\sll$  and the notion of gradation \cite{Kac},
 \cite{al}. As it is well known, $sl(n+1)$ is
 the Lie algebra of the traceless 
$(n+1) \times (n+1)$ matrices.  
 Denote by $E^{ij}$ the elementary 
matrices $E^{ij}=|i><j|,\,\,\, i,j=1\ldots n+1$ 
which satisfy the commutation relations of
 the Lie algebra $gl(n+1)$
\a
& &\left[E^{ij},E^{kl}\right]
=\delta^{jk}E^{il}-\delta^{il}E^{kj}
\label{2.1}
\b   

The Cartan subalgebra $\H$ is generated by 
the traceless combinations of the 
diagonal matrices $E^{ii}$ $H_{\xi}=\sum_{i=1}^{n+1}
\xi_iE^{ii},\,\,\,
\sum_i\xi_i =0$. 
 The rank of $sl(n+1)$ is $n$. To describe the root 
system we fix an orthonormalized basis $\{e_i\}$ in the 
$n+1$--dimensional Euclidean space. Then the  roots
 are exhausted by $\al_{ij}=e_i-e_j,~~ i\neq j$. 
The corresponding step operators are
 $E^{\al_{ij}}=E^{ij}$. As simple roots one can
 choose the elements $\pm \al_i= \pm (e_i-e_{i+1})
,\,\,\, i=1, \ldots n$. The step operators satisfy
 the commutation relations
\a
& &\left[H_\xi,E^{\pm\al_i}\right]=
\pm\al_i.\xi E^{\pm\al_i}=\pm(\xi_i-
\xi_{i+1})E^{\pm\al_i}\0\\
& &\left[E^{\al_i},E^{-\al_j}\right]=
\delta_{ij}H_{\al_i}\label{2.2}
\b
The rest of the step operators is obtained 
by taking successive commutators of the step 
operators  $E^{\al_i}$ and of their
 transposed  $E^{-\al_i}$.
 The highest root  is $\psi = \al_1+...+\al_n = 
e_1 - e_{n+1}$.   
 This  can be translated in the language of the step
 operators: $\left[E^\psi, E^\al\right]=0$ for any 
step operator related to a positive root $\al$. 
Note also that $E^\psi=E^{1n+1}$.
One also introduces the extended root
$\al_0=-\psi$ and its step operator 
 $E^{\al_0}=E^{n+11}$.

We proceed by recalling some facts about the finite 
order inner automorphisms of the simple Lie algebras.
 There is a general theorem due to Kac \cite{Kac} 
(for an introductory review see also \cite{gDS}) which 
states that the finite order inner automorphisms of a 
simple Lie algebra $\G$ are parametrized by $r+1$ 
relatively prime non-negative integers  where $r$ 
stands for the rank of the Lie algebra. In what 
follows, we shall need a special inner automorphism
 $\s$  of $\G=sl(n+1)$, the order of which is 
$n+1,~~\s^{n+1}=1$. To define it, we first recall 
that the fundamental weights $\l_i$
are dual to the simple roots 
$2\frac{\al_i \cdot \l_j}{\al_i\cdot \al_i}=
\delta_{ij},\,\,\, i,j=1,\,\ldots ,\, r$.
 Specifying  $\G=sl(n+1)$ one gets
\a
\l_i&=&\sum_{k=1}^i e_k - \frac{i}{n+1}
\sum_{k=1}^{n+1}e_k\0\\
i, j &=& 1,..,n\label{2.3}
\b
Consider also the vector in the root space
\a
\rho=\sum_{i=1}^n \l_i\label{2.4}
\b  
Using the above notations one  defines the 
following inner automorphism $X \rightarrow \s (X)$ 
of the Lie algebra $sl(n+1)$ \cite{Olive}, 
\cite{Cat}, \cite{Kac}
\a
\s(X)&=&S X S^{-1}\0\\
S&=&e^{ 2\pi i \frac {H_\rho}{n+1}}\label{2.5}
\b
where the diagonal matrix $H_\rho\in \H$ depends linearly 
on the vector $\rho$ (\ref{2.4})
$H_{\rho}=\sum_{k=1}^{n+1} \rho_k |k><k|$; $\rho_k$ 
are the components of (\ref{2.4}) in the basis $\{e_i\}$ 
(\ref{2.3}). In view of (\ref{2.2}), 
$\s$ acts diagonally in the corresponding Cartan--Weyl basis
\a
\sigma(H_\xi)&=&H_\xi\0\\
\sigma(E^{\al_{kl}})&=& \om^{\al_{kl}
\cdot \rho}E^{\al_{kl}}= \om^{l-k}E^{\al_{kl}}\0\\
\om&=&e^{ \frac{ 2\pi i}{n+1}}\label{2.6}
\b
from where it becomes clear that $\sigma$ has order
 $n+1$. Note that the Lie  algebra $\G=sl(n+1)$, 
equipped by the above automorphism becomes a
 graded Lie algebra:
\a
\G&=&\oplus_{k\in{\ZZ}_{n+1}}\G_k\0\\
\sigma(\G_k)&=&\om^k\G_k\0\\
\left[\G_k, \G_l \right]&\subseteq 
&\G_{k+l}\label{2.7}
\b 
in the above equations and in what follows
 we adopt the following convention:
the summation of indices which take values 
in the cyclic group ${\ZZ}_{n+1}$ is 
understood modulo $n+1$.

There exists an alternative basis in 
$sl(n+1)$ closely related to the automorphism
 (\ref{2.5}). We shall briefly review its
 construction (for further details, 
see \cite{Olive}, \cite{Cat}). First of 
all one observes that the generators
\a
\E_i&=& \sum_{k=1}^{n+1-i} E^{kk+i}+
\sum_{k=1}^i E^{n+1+k-i k}=\0\\
&=&\sum_{k\in {\ZZ}_{n+1}}
 E^{k k+i},\,\,\, i=1,\ldots n
\label{2.8}
\b
are mutually commuting. Therefore they span
 another Cartan subalgebra
$\H^{'}$. In the second identity of the above
 equation  the summation index $1\leq k\leq n+1$
 is read modulo $n+1$. Due to (\ref{2.6}), the
 elements (\ref{2.8}) are eigenvectors of
 the automorphism $\s$
\footnote{ Here we only consider the Lie
 algebra $sl(n+1)$. For general simple 
Lie algebras, the eigenvalues of the corresponding
 automorphism, restricted to the alternative Cartan
 subalgebra are related to the Betti numbers}
\a
\s (\E_i)&=&\om^i \E_i
\label{2.9}
\b
Fixing the defining $n+1$--dimensional
 representation, it is not difficult to verify 
that the matrix with entries
\a
\co_{ij}&=& \om^{(i-1)(j-1)}\0\\
\co^{-1}_{ij}&=&\frac{1}{n+1}
\om^{-(i-1)(j-1)}, \,\,\, 1\leq i,j\leq n+1
\label{2.10}
\b
diagonalizes $\H^{'}$
\a
\co^{-1} \E_i \co&=& \sum_{k=1}^{n+1}
 \om^{i(k-1)} E^{kk}
\label{2.11}
\b
From the  expression
\a
S&=& \om^{\frac{n}{2}} \sum_{k=1}^{n+1} \om^{1-k}
 E^{kk}\label{2.12}
\b
for the operator $S$ which implements the 
automorphism (\ref{2.5}), and taking into 
account (\ref{2.11}) we get the commutation
 relations
\a
\co^{\pm 1} \E_k \co^{\mp 1}&=& \om^{\mp \frac{kn}{2}} 
S^{\pm k}\label{2.13}
\b
To complete the alternative basis, we need to 
add the corresponding to 
the Cartan subalgebra $\H^{'}$ step operators. 
In view of (\ref{2.11}) we see 
that 
\a
F^{ij}&=& \co E^{ij} \co^{-1},
\,\,\, i\neq j \label{2.14}
\b
are eigenvectors of the adjoint action of 
the generators (\ref{2.8}). Combining the last 
observation with (\ref{2.9}) we conclude that the 
automorphism (\ref{2.5}) permutes the step 
operators (\ref{2.14}). We calculate explicitly 
its action  with the help of (\ref{2.13})
\a
\s (F^{ij})&=&S\co E^{ij} \co^{-1}S^{-1}=
\co \E_1 E^{ij} \E_n \co^{-1}=\0\\
&=&\co E^{i-1 j-1} \co^{-1}= F^{i-1 j-1}\0\\
i,j&=& 1,\,\ldots ,\, n+1 \, \rm{mod} \,( n+1)
\label{2.15}
\b
It is seen from the above expression that $\s$ 
acts on the alternative basis
as an element of the Weyl group (more precisely, 
it is a Coxeter element). Moreover, it is clear 
that the action of $\s$ separates the set of the 
step operators (\ref{2.14}) into $n$ 
non--intersecting orbits, parametrized by the 
difference $i-j$ (\ref{2.15}), each containing 
$n+1$ elements. Within the elements of each   
$\s$--orbit we choose the following representatives 
\ai
F^{i}&=& \co E^{i+11} \co^{-1}=F^{i+1 1},\,\,\, 
i=1, \ldots n
\label{2.16a}\\
\left[ \E_i , F^j\right]&=& ( \om^{ij}-1)F^j 
\label{2.16b}
\bj
Introducing the grade expansions (\ref{2.7}) of 
the above generators 
\a
F^i&=&\sum_{k\in {\ZZ}_{n+1}} F^i_k\0\\
\s (F^i_k) &=&\om^k F^i_k
\label{2.17}
\b
we observe that the rest of the elements in the 
$\s$--orbit of $F^i$ (\ref{2.16a})
are linear combinations of the components $F^i_k$.
Therefore one gets a graded basis which is formed by
  $\E_i,\, i=1\ldots n$ (\ref{2.8}) and
 $F^i_k,\, i=1,\ldots n;\, k\in {\ZZ}_{n+1}$.
 Note that due to (\ref{2.9}), (\ref{2.16b}) and 
(\ref{2.17}) the commutation relations 
\a
\left[\E_i , F^j_k\right]&=& ( \om^{ij}-1)F^j_{k+i} 
\label{2.18}
\b
are valid.

Starting from a Lie algebra $\G$ one associates 
to it the loop algebra 
$\tilde{\G}={\CC}[\l,\l^{-1}] \otimes \G$, i. e. 
it is the set of the Laurent series on a formal 
parameter $\l$, which will play the role of a 
spectral parameter of the auxiliary linear
 problem, with coefficients belonging to $\G$.
In other words, $\tilde{\G}$ is spanned on the
 elements $X_n=\l^nX$ where
$n\in {\ZZ}$ and $X\in \G$. The commutator is
\a
\left[X_m , Y_n\right]&=& \left[X,Y\right]_{m+n} \0
\b
Now we introduce the the Lie algebra $\sll$ in 
the principal gradation. It is generated by 
the expansions
$X(\l)=\sum_{l\in{\ZZ}} \l^l X_l$ with $X_l\in sl(n+1)$ 
together with the 
restriction 
\a
X(\om \l)&=& \s ( X(\l))\label{2.19}
\b
where the automorphism $\s$ acts on the Laurent
 coefficients $X_n$ as indicated by (\ref{2.5}). 
The operator $d=\l \frac{\del}{\del \l}$ 
introduces a ${\ZZ}$--gradation in $\tilde{\G}=\sll$ 
\a
\widetilde{\G}=\oplus_{n\in\ZZ}~\widetilde{\G_{n}}\0\\
\left[d, \widetilde{\G}\right]=n\widetilde{\G_n}
\label{2.20}
\b
Comparing (\ref{2.7}) with (\ref{2.19}) and taking 
into account the above decomposition one concludes 
that $\G_k  \simeq \tilde{\G}_{k+l(n+1)}$ for
 $k\in {\ZZ}_{n+1}$ and $l \in \ZZ$. Starting from
  the alternative basis (\ref{2.8}), 
(\ref{2.17})  on $sl(n+1)$ one gets a basis of $\sll$ 
in the principal gradation. It is formed by the elements
$\E_{i+l(n+1)}$ for $i=1,\ldots n,\, l\in \ZZ$ and
 $F^j_{i+l(n+1)}$ for
$j=1,\ldots n,\, i\in {\ZZ}_{n+1},\, l\in \ZZ$.
 The subalgebra generated by 
$\E_k$, $k\neq 0\, \,{\rm mod}\, (n+1)$ is a
 maximal abelian subalgebra. It is known in the
 literature as the principal Heisenberg 
subalgebra\footnote{ The Heisenberg subalgebras of the loop 
and the affine Lie algebras play crucial role in 
constructing integrable hierarchies \cite{DS}--\cite{eng}}. 
Introducing the element \cite{Olive}, \cite{Cat}
\a
F^i(\mu)&=&\sum_{l\in \ZZ} \mu^{-l} F^i_l\label{2.21}
\b
and taking into account the commutator (\ref{2.18}) 
and its extension in the loop algebra, 
one obtains\footnote{Here and in what follows 
we will perform a slight abuse of notations, namely 
the lower index will be used to indicate both the
 discrete ${\ZZ}_{n+1}$ and to parametrize the
 ${\ZZ}$ gradation as
introduced in (\ref{2.20})}
\a
[\E_i , F^j(\mu) ] &=& (\om^{ij}-1) \mu^i F^j(\mu)
\label{2.22}
\b

To introduce the $A_n^{(1)}$ Toda equations we
 first define the following element of 
the Cartan subalgebra 
\a
\Phi=\frac{1}{2}\sum_{i=1}^{n+1}
\var_i E^{ii}~~~~~~~\sum_i\var_i=0\label{2.23}
\b
Then the $A_n^{(1)}$ \aTe\, can be 
written in the following form:
\a
\del_+\del_-\Phi&=&m^2\left[e^{ad\Phi}
\E_+, e^{-ad\Phi}\E_-\right]\0\\
x^{\pm}&=&x\pm t,~~~~~~ \del_{\pm}=
\frac{\del}{\del x^\pm}\0\\
& &e^{adX}Y=e^X.Y.e^{-X}\label{2.24} 
\b
where $\E_{\pm}$ are grade $\pm 1$ elements 
of the principal Heisenberg subalgebra of 
$\sll$. More precisely, they are liftings of 
$\E_1$ and $\E_n$ (\ref{2.8}) in the loop algebra
\a
\E_{\pm}&=& \l^{\pm 1}\sum_{k\in {\ZZ}_{n+1}}
 E^{k k\pm 1}\label{Heis}
\b
Substituting back the above expressions into
 the equations of motion (\ref{2.24})
and taking into account the notation 
(\ref{2.23}) we end up with the system
\a
\del_+\del_-\var_i&=&m^2(e^{\var_i-\var_{i+1}}
-e^{\var_{i-1}-\var_i})\0\\
i&\in & {\ZZ}_{n+1}\label{2.25}
\b
It is an easy task to check that (\ref{2.24})
 is equivalent to the zero-curvature condition
\a
\del_+A_--\del_-A_++\left[A_+, A_-\right]=0\label{2.26}
\b
of a connection whose components belong to the  loop
 algebra $\sll$ in the principal gradation:
\a
A_+(x,\l)&=&2\del_+\Phi(x)+m\E_+\0\\
A_-(x,\l)&=&me^{-2ad\Phi(x)}\E_-\0\\
\label{2.27}
\b
where we adopted the abbreviation $x=(x^+, x^-)$.
The dependence on the spectral parameter in the
 above expressions comes from the dependence of
 the elements of the principal Heisenberg
 subalgebra $\E_\pm$ on it. The zero curvature
 condition (\ref{2.26})
implies that there exists covariantly constant
 vector $w( x, \l)$ with respect to the covariant
 derivatives $D_{\pm}=\del_{\pm}+A_{\pm}$:
\a
(\del_{\pm}+A_{\pm}(x,\l))w(x,\l)=0\label{2.28}
\b
In what follows we shall assume that the above
 equation is considered in the defining 
representation of $\sl$. Since the components
 $A_{\pm}$ are in the
principal loop algebra $\sll$, they obey the
 relations (\ref{2.19}). Performing the rescaling
 of the spectral parameter
 $\l \rightarrow \om^{-1} \l$,
 one immediately observes that (\ref{2.28})
 remains invariant provided that 
\a
w(x,\l) &\rightarrow & (\S w)(x,\l)\0\\
(\S w)(x,\l)&=& S w(x,\om^{-1} \l) \label{2.29}
\b
where the matrix $S$ implements the automorphism 
(\ref{2.5}). The above symmetry of the equation 
(\ref{2.28}) allows to get a matrix 
solution of it, starting 
from the column $w(x,\l)$
\a
W(x,\l)&=&\parallel w(x,\l),
\om^{\frac{n}{2}}(\S^{-1}w)(x,\l)
\ldots 
\om^{\frac{n^2}{2}}(\S^{-n}w)(x,\l)\parallel 
\label{2.30}
\b
Note that the last expression justifies our choise 
to work with the defining representation. Since the 
order of the automorphism $\s$ (\ref{2.5}) is $n+1$
(\ref{2.6}), it is clear that the $n+1$--th power
 of the operator $S$ (and therefore of $\S$
 (\ref{2.29})) which implements this automophism is
 proportional to the identity operator in any 
irreducible representation of $\sl$. This restricts 
us to look for matrix solutions of (\ref{2.28}), 
starting from a vector one, by using the symmetry 
(\ref{2.9}) in the defining representation only 
since
 it  has dimension $n+1$.

To get soliton solutions we shall look for special 
solutions of the system (\ref{2.28}) which admit 
the expansion
\a
w(x,\l)&=&\sum_{j=0}^N \l^jw^{(j)}(x)
e(x,-\l)\0\\
e(x,\l)&=&exp \{m(\l x^++\frac{x^-}{\l})\}
\label{2.31}
\b  
where N is non--negative integer which will
 be identified with the number of  solitons 
and $w^{(j)}(x), ~j= 1,\,\ldots ,\, N$ are 
$\l$--independent  vectors. We shall require
 also that $w^{(N)}$ is the constant vector 
with unit components
\a
w^{(N)}&=&\sum_{k=1}^{n+1}|k\rangle\0\\
\E_{\pm}w^{(N)}&=&\l^{\pm1}w^{(N)}
\label{2.32}
\b

To fix the rest of the coefficients in the expansion 
(\ref{2.31}) we shall impose the following relations
 on $w(x,\l)$ \cite{Date} 
\a
(\S^{-r_j}w)(x,\mu_j)&=&\om^{-\frac{r_jn}{2}}
c_jw(x,\mu_j)\0\\
j&=& 1,\, \ldots ,\, N\label{2.33}
\b
for certain constants $c_j$ and $\mu_j$ and 
 discrete parameters, called soliton species, 
$r_j$ which take {\it non--vanishing} values in 
the cyclic group ${\ZZ}_{n+1}$. Taking into
 account (\ref{2.12}) and (\ref{2.29}) 
 we conclude that the above equations can be
 equivalently written as follows
\a
w_k(x, \om^{r_j}\mu_j)&=&\om^{(1-k)r_j}
c_jw_k(x,\mu_j)\label{2.34}
\b
where $w_k$ are the components of the vector $w$
\a
w&=&\sum_{k=1}^{n+1} w_k |k>\0
\b
From (\ref{2.33}) we see that the matrix 
$W(x,\l)$ (\ref{2.30})
is degenerate for $\l=\om^k \mu_j$ where 
$k\in {\ZZ}_{n+1}$ and $j=1,\, \ldots ,\, N$. 
For these values of the spectral parameter,
 the columns with numbers $1-k$ mod ($n+1$)
 and $r_j+1-k$ mod ($n+1$) are proportional.

To demonstrate that the expansions (\ref{2.31})
 together with  (\ref{2.32}) and (\ref{2.33})
 satisfy (\ref{2.28}), we shall make the 
following observation: suppose that
 $U(x,\l)=P_{N-1}(x,\l)e(x,-\l)$, where
 $P_{N-1}(\l)$ is a vector valued polynomial
 on $\l$ of degree $N-1$, is a solution of  
 (\ref{2.33}). 
Then $U(x,\l)$ vanishes identically.
 To see this, note that due to (\ref{2.33}),
 the coefficients of the polynomial $P_{N-1}$ 
satisfy certain homogeneous linear system of
 $N$ equations. For generic values of $\mu_j$,
 the determinant of this system is different
 from zero. Therefore there only exist a
 vanishing solution,
and hence $U(x,\l)\equiv 0$. Let us apply
 this observation to 
\a
U_+(x,\l)&=&\del_+w(x,\l)+2\del_+\Phi(x) 
w(x,\l)+m\E_+w(x,\l)\0\\
U_-(x,\l)&=&\del_-w(x,\l)+me^{-2ad\Phi (x)}
\E_-w(x,\l)\label{2.35}
\b
 As consequence of (\ref{2.5}), (\ref{2.6}), 
(\ref{2.9}) and since $w(x,\l)$ is a solution
 of (\ref{2.33}) one checks immediately that 
the above expressions  satisfy (\ref{2.33}) also. 
Inserting (\ref{2.31}) into (\ref{2.35}) we get:
\a
& &e(x,\l)U_+(x,\l)=
\l^N\left(2\del_+\Phi w^{(N)}-
m(1-\frac{1}{\l}\E_+)w^{(N-1)}\right)+
R_{N-1}(x,\l)\0\\
& &e(x,\l)U_-(x,\l)=\frac{m}{\l}
\left(\l e^{-2ad\Phi}\E_--1\right)w^{(0)}(x)
+S_{N-1}(x,\l)\0
\b
where $R_{N-1}$ and $S_{N-1}$ are polynomials
 on $\l$ of degree not greater than $N-1$.
 To derive the first of the above expansions 
we have also used (\ref{2.32}). Therefore 
$U_{\pm}$ vanish identically  provided that
\a
2\del_+\Phi w^{(N)}&=&m\left(1-\frac{1}{\l}\E_+\right)
w^{(N-1)}\0\\
\l e^{-2ad\Phi}\E_-w^{(0)}(x)&=&w^{(0)}(x)\label{2.36}
\b
This method to construct \aTo\,solitons is a 
straightforward generalization of the  approach applied 
by Date  \cite{Date} for the \sG\, model.
Taking into account (\ref{2.8}) and (\ref{2.23}) 
one can write the above expressions as
\a
\del_+\var_i&=&m(w^{(N-1)}_i-w^{(N-1)}_{i+1})\0\\
e^{2\al_i . \Phi}&=& e^{\var_i-\var_{i+1}}=
\frac{w^{(0)}_{i+1}}{w^{(0)}_{i}},\,\,\, 
i\in {\ZZ}_{n+1}\label{2.37}
\b    
To find explicit expressions from (\ref{2.37})
 for the fields $\var_i$ one has   to solve the
 linear equations (\ref{2.34})  which can be 
written as follows:
\a
& &\sum_{l=1}^N \mu_j^{l-1}\left(\om_j^{l-1}
e(-\om_j\mu_j)-c_j\om_j^{1-k}e(-\mu_j)\right)
w_k^{(l-1)}=\0\\
& &=\mu_j^N(c_j\om_j^{1-k}e(-\mu_j)-\om_j^N 
e(-\om_j\mu_j))\0\\
& &\om_j=\om^{r_j},~~~~ k= 1,...,n+1\label{2.38}
\b
where  $w_k^{(l-1)}$ are the components of the
 coefficients $w^{(l-1)}$ which appeared in the
 expansion (\ref{2.31}); here and in what
 follows we shall omit the $x^{\pm}$
 dependence in the exponentials $e(x,\l)$
 (\ref{2.32}). By Kramer's formula we get
 the solution
\a
w^{(0)}_{k}&=&(-)^N \prod_{j=1}^{N}\mu_j
 \om_j\frac{{\rm det}~G^{(k+1)}}
{{\rm det}~G^{(k)}}\0\\
w_k^{(N-1)}&=&\frac{1}{m}\del_+
\ln~{\rm det}~G^{(k)},\,\,\,
k=1,\ldots,n+1\label{2.39}
\b 
where $G^{(k)}$ is a $N\times N$ matrix with 
entries 
\a
G_{jl}^{(k)}&=&\mu_j^{l-1}(\om_j^{l-1}e(-\om_j\mu_j)
-c_j\om_j^{1-k}e(-\mu_j))
\label{2.40}
\b
There is a "canonical" expression for the
 determinant of $G^{(k)}$ \cite{Date}. A way
 to obtain it is to multiply and divide by the
 Van der Mond determinant det$M$ where
 $M_{jl}=\mu_j^{l-1}$ ($j,l=1,...,N$), and to
 calculate the matrix elements of product
 $G^{(k)}M^{-1}$. The final result reads 
\a
{\rm det}\,G^{(k)}&=&(-)^N 
\frac{\prod_{j=1}^Nc_j\om_j^{1-k}e(-\mu_j)}
{\prod_{p>q}(\mu_p-\mu_q)}
\tau_{k-1}\0\\
\tau_k &=& {\rm det}\,
(1+\co^{\frac{k}{2}}.V.\co^{\frac{k}{2}})\0\\
\co&=&{\rm diag}\,(\om_1,..., \om_N)\label{2.41}
\b
where the elements of the matrix $V$ are given by:
\a
& &V_{jk}=\frac{\sqrt{X_j~X_k}}{\mu_j^+-\mu_k^-}\0\\
& &X_j=\frac{1}{c_j}(\mu_j^--\mu_j^+)\prod_{l\neq j}
\frac{\mu_l^--\mu_j^+}{\mu_l^--\mu_j^-}
{\rm exp}\{-2im\sin\frac{\pi r_j}{n+1}
(\widetilde{\mu_j}~x^+-\frac{x^-}
{\widetilde{\mu_j}})\}\0\\
& &\widetilde{\mu_j}=\om_j^{\frac{1}{2}}
\mu_j,~~~~\mu_j^{\pm}=\om_j^{\pm\frac{1}{2}}
\widetilde{\mu_j}\label{2.42}
\b
Substituting back (\ref{2.39}) into (\ref{2.37})
 we get:
\a
e^{-\var_k}=\frac{\tau_k}{\tau_{k-1}}=
\frac{{\rm det}\,( 1+\Omega^{\frac{k}{2}}.V.
\Omega^{\frac{k}{2}})}
{{\rm det}\,(1+\Omega^{\frac{k-1}{2}}.V.
\Omega^{\frac{k-1}{2}})}
\label{2.43}
\b
The  above expression for the\aTo\, N--solitons
 permits to establish a relation to the relativistically
 invariant N--body integrable systems of Calogero--Moser type
\cite{body}. 

From (\ref{2.41}), (\ref{2.42}) it is seen that the solutions 
(\ref{2.43}) describe propagation of $N$ solitons: 
the variables $\widetilde{\mu_j}$ are the rapidities while
the quantities 
\a
a_j=\frac{1}{c_j}(\mu_j^--\mu_j^+)
\prod_{l\neq j}\frac{\mu^-_l-\mu^+_j}
{\mu^-_l-\mu^-_j}~~,\0\\
\b
are related to the positions. Note that more 
than the continuous parameters $a_j$ and $\mu_j$, 
the solitons are characterized by the discrete
 parameters $r_j$ (\ref{2.33}). As a final remark of
this Section, we observe that the solutions with 
oscillations around the
standard solitons which has been studied recently
in details \cite{BJ}, appear as particular cases of the general
expressions (\ref{2.41})--(\ref{2.43}) for particular
values of the rapidity--type parameters $\mu_j$.

\section{ The dressing problem for the $\An$ Toda solitons}

\setcounter{equation}{0}
\setcounter{footnote}{0}
The dressing group is a symmetry of the
 non--linear evolution equations which admit zero--curvature
 (or Lax) representation. It was shown in \cite{fr} that 
the \dg\, appears as a semiclassical limit of the quantum
 group symmetry. The \dg\, acts by gauge transformations
 on the components of the Lax connection which preserve its
 form. Therefore, it is a symmetry of the space of solutions
 of the corresponding integrable model. The aim of this
 Section is to present a derivation of the \dg\, elements
 which generate $N$--solitons in the $\An$ Toda theories.
 This problem has  been already solved in \cite{BB} for 
the \sG\, equation
which is the $A_1^{(1)}$ Toda model. Expressions for
 the \dg\, elements which 
create solitons from the vacuum have been conjectured 
in \cite{ago} for a large class of integrable hierarchies. 

In order to be able to compare our results with the
 expressions of Babelon 
and Bernard \cite{BB}, it will be convenient to
 perform a field dependent gauge transformation on the
 components of the connection (\ref{2.27}) 
\a
D_{\pm}&\rightarrow & e^{\Phi} D_{\pm} e^{-\Phi}=
 \del_{\pm}+\A_{\pm}\0\\
\A_{\pm}&=& \pm \del_{\pm} \Phi +m e^{\pm \rm{ad} \Phi}
 \E_{\pm}\label{3.1}
\b
It is clear that for the vacuum solution $\Phi=0$, the
 components of the above connection take the following
 form
\a
\A_{\pm}&=& m \E_{\pm}  \label{3.2}
\b
In accordance with the general definition, the
 \dts\, are represented by loop group elements
  $g(x,\l) \in \Sll$ which act on (\ref{3.1})
 as gauge transformations $\A_{\pm} 
\rightarrow \A_{\pm}^g$
\a
\A_{\pm}^g&=& - \del_{\pm} g g^{-1}+g\A_{\pm}g^{-1}
\label{3.3}
\b
such that the connection $\A_{\pm}^g$ has  the same 
form as the original one
(\ref{3.1}) with $\Phi \rightarrow \Phi^g$. 
Since by gauge transformations the curvature
 transforms as $F_{+-}=[D_+ ,D_-] 
\rightarrow g F_{+-} g^{-1}$ we see that  the
 \dts\, are symmetries of the underlying equations
 of motion (\ref{2.25}), (\ref{2.26}).

In view of (\ref{3.1}), it is clear that 
\a
\T(x,\l)&=& e^{\Phi(x)} W(x,\l)\label{3.4}
\b
where the matrix $W$ was introduced by 
(\ref{2.30}), is a solution of the linear 
problem
\a
\left( \del_{\pm} + \A_{\pm} \right)\T(x,\l)
 &=&0 \label{3.5}
\b

Due to the expansion (\ref{2.31}), it is obvious
 that the components $w_k$ of the 
($n+1$--dimensional) vector $w$ admit the
 following representation
\a
w_k(x,\l)&=& \prod_{j=1}^N (\l+\ep_{kj}(x)) 
e(x,-\l),~~~~1\leq k \leq n+1 \label{3.6}
\b
The dependence of the variables $\ep_{kj}$ 
on $x^+$ and $x^-$ is fixed by 
(\ref{2.34}) which, taking into account the 
above expression, reads
\a
\prod_{l=1}^N \frac{\ep_{kl}+\o^{r_j}\mu_j}
{\ep_{kl}+\mu_j}&=&c_j \o^{r_j(1-k)}
\frac{e(\o^{r_j}\mu_j)}{e(\mu_j)}\label{3.7}
\b
In view of (\ref{3.6}) we can express
 the matrix (\ref{2.30}) as follows
\ai
W(x,\l)&=& U(x,\l) E(x,\l) \label{3.8a}\\
U_{kl}(x,\l)&=& \o^{(k-1)(l-1)}
 \prod_{j=1}^N(\ep_{kj}+\o^{l-1}\l
)\label{3.8b}\\
E_{kl}(x,\l)&=&\delta_{kl}
 e(-\o^{k-1}\l),~~~~~~k,l=1,\ldots, n+1
\label{3.8c}
\bj

As a next step, we shall calculate the
 determinant of (\ref{2.30}). First of
 all we note that due to the above equations,
 the exponential singularities of the matrix
 elements of $W$ disappear in its determinant.
 Therefore, $\rm{det}\, W$ is a meromorphic
 function on the 
Riemann sphere ${\CC} {\PP}^1$. Further, due to
 (\ref{2.34}) $\rm{det}\, W$
vanishes  whenever $\l^{n+1}=\mu_j^{n+1}$
 for $j=1,\ldots ,N$. This wants to say
 that $\rm{det}\, W$ has  at least $N(n+1)$
 zeroes. This number is exact since one has 
 the expansion
\a
\rm{det}\, W(x,\l)&=&(-)^{nN}\l^{(n+1)N}
\rm{det}\, 
\co \left(1+O(\frac{1}{\l})\right),~~~
\l\rightarrow \infty 
\label{3.9}
\b
where the matrix $\co$ is given by (\ref{2.10}). 
It is clear that $\rm{det}\, W$
has  no other poles, and therefore, due to 
the Cauchy's theorem we end up with the result
\a
\rm{det}\, W(x,\l)&=& (-)^{nN}\rm{det}\, \co 
\prod_{j=1}^N (\l^{n+1}-\mu_j^{n+1}) \label{3.10}
\b
In what follows it will be necessary to express 
the $\An$ Toda fields (\ref{2.23}), (\ref{2.25})
 in terms of the variables\footnote{These
 variables appeared in the study of the 
periodic solutions of the KdV equation and
 of the periodic Toda chain\cite{FMc}}
 $\ep_{kl}$ (\ref{3.6}). To do that, it suffices 
to compare (\ref{2.31}) with (\ref{3.6}). 
The result is
\a
w_k^{(0)}&=&\prod_{j=1}^N \ep_{kj}\label{3.11}
\b
which together with (\ref{2.37}) and 
(\ref{2.39}) yields the expression
\a
e^{-\var_k}&=& (-)^N \prod_{j=1}^N
\frac{\ep_{kj}}{\mu_j},~~~~k=1,\ldots , n+1
\label{3.12}
\b
Since the field $\Phi$ (\ref{2.23}) belongs to 
the Cartan subalgebra of $\sl$,
the following restriction
\a
\prod_{k=1}^{n+1} \prod_{j=1}^N 
\ep_{kj}&=&(-)^{N(n+1)}  \prod_{j=1}^N 
\mu_j^{n+1}\label{3.13}
\b
takes place.

Turning back to the dressing problem, 
we define the normalized transport 
matrix\footnote{As a reference point we 
choose those with light--cone coordinates 
$x^+=x^-=0$}
\a
T(x,\l)&=&\T(x,\l) \T^{-1}(0,\l)\label{3.14}
\b
It is obvious that the above matrix 
is unimodular. Moreover, due to (\ref{3.5}), 
it belongs to the \lg\, $\Sll$ in 
the principal gradation. Let $\T$ (\ref{3.4})
and $T$ (\ref{3.14}) be the transport 
matrices associated to certain
$N$--soliton solution (\ref{2.43}), 
(\ref{3.12}) and $\T_0$, $T_0$ be the
 corresponding vacuum transport matrices.
 Therefore one can write
\ai
T(x,\l)&=& f(x,\l) T_0(x,\l) f^{-1}(0,\l) 
\label{3.15a}\\
f(x,\l)&=& \T(x,\l) \T_0^{-1}(x,\l)=
e^{\Phi(x)} W(x,\l) W_0^{-1}(x,\l)=\0\\
&=&e^{\Phi(x)}U(x,\l)U_0^{-1}(x,\l)
\label{3.15b}\\
T_0(x,\l)&=&\T_0(x,\l)
\T_0^{-1}(0,\l)=e^{-m(\E_+x^++\E_-x^-)}
\label{3.15c}
\bj
where $W(x,\l)$ and $W_0(x,\l)$ are the 
matices (\ref{2.30}) corresponding to a 
generic $N$--soliton solution (\ref{2.43})
 and to the vacuum respectively; the 
matrix $U$ in
 the last equation (\ref{3.15b}) 
is given by (\ref{3.8a})--(\ref{3.8c}) while
 $U_0$ stands for its vacuum solution analogue.
Due to (\ref{3.8b}) it turns out that $U_0=\co$
(\ref{2.10}). Taking into account these remarks
 we conclude that $f(x,\l)=e^{\Phi(x)} U(x,\l)\co^{-1}$. 
In view of (\ref{3.15a}), $f(x,\l)$ is a gauge
transformation which transforms the vacuum solution 
transport matrix $T_0$ into the transport matrix $T$,
 related to a $N$--soliton solution. Therefore, it
 generates a \dt . Note also that the element
 $f(x,\l)$ is in the principal gradation
\a
f(x,\o \l)&=&S f(x,\l) S^{-1} \label{3.16}
\b
To get the above equation we observe that 
the matrix (\ref{3.8b}) satisfies
 the equation $U(x,\o \l)=\l
 \o^{-\frac{n}{2}} U(x,\l)\E_-$ 
which combined with the commutation
 relations (\ref{2.13}) produces 
(\ref{3.16}). Due to (\ref{2.19}) we see
 that $f(x,\l)$ belongs to the loop group
 $\widetilde{GL}(n+1)$ in the principal
 gradation. From (\ref{3.10}) it is seen that
\a
\rm{det}\, f(x,\l)&=&(-)^{nN}
\prod_{j=1}^N(\l^{n+1}-\mu_j^{n+1})\label{3.17}
\b
Note that the solution of the dressing problem
 (\ref{3.15a}) is not unique. 
The reason is that there exist
 $x^{\pm}$--independent
 matrices $\th (\l)$ 
which are not proportional to the identity
 and commute with $T_0$ (\ref{3.15c}).
Therefore the element
\a
g(x,\l)&=&f(x,\l) \th (\l) \label{3.18}
\b
is the general solution of the dressing problem;
 in the above equation $f(x,\l)$ (\ref{3.15b})
 is a particular solution of it. To fix the
 unknown matrix $\th (\l)$ we impose a set
of restrictions
\ai
\th(\l) \E_{\pm}\th^{-1}(\l)&=&
 \E_{\pm}\label{3.19a}\\
\th(\o \l)&=& S \th(\l) S^{-1}
\label{3.19b}\\
\rm{det} \th(\l)&=& 
\frac{(-)^N}{\prod_{j=1}^N(\l^{n+1}-\mu_j^{n+1})}
\label{3.19c}\\
&&\\
g(x,\l)&=&\left\{\begin{array}
{ccc}
e^{-\Phi}\left(1+ O(\l)\right)& 
\rm{for}& \l\rightarrow 0\\

\\

e^{\Phi}\left(1+ O(\frac{1}{\l})\right)
&\rm{for}& \l\rightarrow \infty 
\end{array} \right. \label{3.19d}
\bj
To justify the above requirements we
 note that (\ref{3.19a}) ensures the
 commutativity of $\th(\l)$ with $T_0$ 
(\ref{3.15c}); taking into account 
(\ref{3.16}), we see that (\ref{3.19b}) 
 guarantees that the \dg\, element $g(x,\l)$ 
is in the principal gradation; (\ref{3.19c}) 
comes from the requirement that $g(x,\l)$
 should be unimodular; finally, (\ref{3.19d})
 is a consequence of a grade analysis
 applied to (\ref{3.3}) 
(for details, see \cite{fr}, \cite{BB}). 
It is easy to check that the general 
solution (\ref{3.19b}) is given by
\a
\th (\l)&=& \th_0(\l) + \sum_{k=1}^n 
\th_k(\l)\E_k\label{3.20}
\b
where the generators $\E_k$ of the 
alternative Cartan subalgebra $\H^{'}$ 
were introduced by (\ref{2.8}). Inserting 
the above expansion into (\ref{3.19b}) 
and taking into account (\ref{2.9})
 we end up with
\a
\th_k (\o \l)&=& \o^k 
\th_k(\l),~~~k=0,\ldots,n
\label{3.21}
\b
To calculate the determinant
of $\th(\l)$ we shall use (\ref{2.11})
\a
& &\rm{det}\, \th(\l)= \rm{det}\, \co^{-1} 
\th (\l) \co=\prod_{k=1}^{n+1}
\left( \sum_{l=1}^{n+1} \co_{kl} 
\th_{l-1}(\l)\right)=\0\\
& &= \prod_{k=1}^{n+1}\left( \sum_{l=1}^{n+1} 
\th_{l-1} (\o^{k-1}\l)\right)
\label{3.22}
\b 
More than (\ref{3.19a})--(\ref{3.19d}),
 we shall require that the entries of 
the matrix (\ref{3.18}) are meromorphic
functions on ${\CC}{\PP}^1$ with simple
 poles located at the points 
$\l=\o^p \mu_j$ for
 $p=0,\ldots , n$ and 
$j=1, \ldots , N$. In view of the last restriction,
 only a finite number of solutions survive. Among
 them we choose that which satisfies the system
\a
\sum_{l=1}^{n+1} \co_{kl} \th_{l-1}(\l)&=&
 \frac{1}{\prod_{j=1}^N
(\o^{k-1}\l-\mu_j)},~~~ k=1,\ldots , n+1 
\label{3.23}
\b
It is clear that (\ref{3.22}) together with
 the above equation guarantee the the
 validity of (\ref{3.19c}). Note also that 
(\ref{2.23}) is compatible with
 (\ref{3.21}). Inserting back (\ref{3.20}),
 (\ref{3.23}) into (\ref{3.18}) we obtain 
\ai
g^{(N)}(\Phi,\{\mu\},\l)&=& e^{\Phi}
\Gamma^{(N)}(\Phi,\{\mu\},\l) \co^{-1}
\label{3.24a}
\b
where the upper index indicates the number of
 solitons and
 $\Gamma^{(N)}$ is a $(n+1)\times(n+1)$ matrix
 with entries
\a
\Gamma_{kl}^{(N)}(\Phi,\{\mu\},\l)&=&
 \o^{(k-1)(l-1)} \prod_{j=1}^N 
\frac{\l+\o^{1-l}\ep_{kj}(x)}{\l-\o^{1-l}\mu_j}
\label{3.24b}
\bj
Note that the dependence on the space--time
 coordinates is dictated by (\ref{3.7}) and
 (\ref{3.12}).
The expansion (\ref{3.19d}) is satisfied as
 a consequence from (\ref{3.12}). We note
 that the method presented in this Section 
was previously used in \cite{R}
to solve the dressing problem for the 
algebraic--geometrical solutions in the \sG\, model.

\section{ The factorization problem and the relation
 to the vertex operator approach}

\setcounter{equation}{0}
\setcounter{footnote}{0}

There exists a general scheme to construct solitons 
in the \aTo\, theories
\cite{Olive}. Substantially, it is based on the
 group--algebraic approach 
to the integrable systems, developed by Leznov
 and Saveliev \cite{LS}.
To apply the Leznov--Saveliev analysis to the \aTo\,
 equations one first considers the \CaT\, (CaT)
 equations \cite{cat},\cite{SP}. The last appear as a
 zero--curvature condition of a connection
 of the form (\ref{2.27}) the components of
 which belong to the affine Lie algebra
 $\hat{\G}$. It is the central extension
 of the corresponding loop algebra $\tilde{\G}$.
The necessity to introduce central extension of
the loop algebra is due to the fact that the 
Leznov--Saveliev analysis  applies to Lie
algebras which admit (non--trivial) highest
weight representations. Such representations
only exist if the central charge is different
from zero.
 In the case of the CaT models, the
 group--algebraic approach yields the
 general solution of the equations of
 motion, parametrized by a free massless
 field and a group element 
which belongs to the affine Lie group
 $\hat{G}$. It was suggested in \cite{Olive}
 that solitons arise when the group element
 factorizes in a product of special elements
of the affine Lie group which are closely related
to the vertex operators.
 These elements are exponentials of the loop
 algebra elements which diagonalize the adjoint
 action of the principal Heisenberg
 subalgebra. Within 
the formalism of \cite{Olive}, insertion of
 one such element results in a
creation of a single soliton. The
 group--algebraic approach to the solitons 
in the \aTo\, theory was further developed
 in \cite{BB}. In the last paper,
on the example of the conformally extended 
\shG\, model, it was shown that the solitons
 can be obtained from the vacuum via specific
 dressing transformations.
The explicit form of the corresponding
 dressing group elements has  been used 
to establish a relation to the vertex
 operator formalism \cite{Olive}. Moreover,
 it was demonstrated that the solution of
 the dressing problem in the affine group
 differ from those in the loop group by a
 factor which is in the center
\footnote{The arguments used in \cite{BB} 
can be easily
 generalized
to apply to an arbitrary \aTo\, theory}.
 In the present Section we extend the
results of \cite{BB} for the $\An$ Toda models, 
i. e. starting from (\ref{3.24a}), (\ref{3.24b})
 we first show that one can factorize a generic 
dressing group elements into a product of 
"monosoliton" factors; second, we analyze 
our expressions (\ref{3.24a}), (\ref{3.24b})
 for $N=1$ and obtain the relation to the 
vertex operator construction of the soliton
 solutions \cite{Olive}.

We start by writing the element  
(\ref{3.24a}), (\ref{3.24b}) in a 
slightly different form
\ai
g^{(N)}(\Phi,\{\mu\},\l)&=&
\frac{1}{n+1} \sum_{r\in {\ZZ}_{n+1}}
S^r |v_{\Phi}( \{\mu\},\o^{-r}\l)>
<v_0|S^{-r}\label{4.1a}\\
|v_{\Phi}( \{\mu\},\l)>&=&\sum_{i=1}^{n+1}
 e^{\frac{\var_i}{2}}
\prod_{a=1}^N \frac{\l+\ep_{ia}}{\l-\mu_a}|i>\0\\
|v_0>&=&\sum_{i=1}^{n+1} |i>\label{4.1b}
\bj
where the operator $S$ was introduced by 
(\ref{2.5}), (\ref{2.12}). It is clear that
the property (\ref{3.16}) is manifestly 
satisfied by the expression (\ref{4.1a}). 
Note that the vector $|v_0>$ already
 appeared in a different context (\ref{2.32}).
 We proceed by advancing the hypothesis that
 (\ref{4.1a}) 
admits the representation
\a
g^{(N)}(\Phi,\{\mu\},\l)&=&e^{\P_N} 
g^{(1)}(F_N,\mu_N,\l)
\cdot\, \ldots \, \cdot e^{\P_1} 
g^{(1)}(F_1,\mu_1,\l) 
\label{4.2}
\b
where
\a
&&\P_l=\frac{1}{2}\sum_{k=1}^{n+1}
p_{kl}E^{kk},~~~~~~~
F_l=\frac{1}{2}\sum_{k=1}^{n+1}f_{kl}
E^{kk}\0\\
&&\sum_{k=1}^{n+1}p_{kl}=
\sum_{k=1}^{n+1}f_{kl}=0,~~~~~l=1, \ldots \, , N
\label{4.3}
\b
are certain elements of the Cartan 
subalgebra of $\sl$; $ g^{(1)}(F_l,\mu_l,\l)$
are monosolitonic factors. They have the same
 form as (\ref{4.1a}), (\ref{4.1b})
with $|v_{\Phi}>$ substituted by
\a
|v_{F_l}( \mu_l,\l)>&=& e^{F_l}\sum_{i=1}^{n+1}
 \frac{\l-\mu_le^{-f_{il}}}{\l-\mu_l}|i>
\label{4.4}
\b
Note that substituting back the above
 expression into (\ref{4.1a})
and taking into account (\ref{3.12}) we 
reproduce (\ref{3.24a}), (\ref{3.24b})
with $N=1$. It is not difficult to calculate
 the inverse element 
\a
\left(g^{(1)}\right)^{-1}(F,\mu,\l)
&=&e^{K(F)} g^{(1)}(-F_,\mu,\l)
 e^{-K(F)}\0\\
K(F)&=&\frac{1}{2}\sum_{i=1}^{n+1}
(H_{\l_i}+H_{\l_{i-1}})f_i
\label{4.5}
\b
where $\l_i$, $i=1, \ldots , n$ are 
the fundamental weights (\ref{2.3}) 
of $\sl$; $\l_0=\l_{n+1}=0$ and
 $\l \rightarrow H_{\l}$ is the
 natural identification of the 
$n+1$--dimensional Euclidean space with the 
$(n+1)\times (n+1)$ diagonal matrices 
$H_{\l}=\sum_i\l_i E^{ii}$. Therefore, the 
last of the above equations can be equivalently 
written as
\a
K_i(F)-K_{i+1}(F)&=&\frac{f_i+f_{i+1}}{2}
\label{4.6}
\b
Note also that $K_i(F)=K_{i+n+1}(F)$ since 
$\sum_i f_i=0$.

To demonstrate the validity of the
 factorized expression (\ref{4.2}) we first 
introduce the notation
\a 
g^{(N)}_l(\Phi,\{\mu\},\l)
&=&g^{(N)}(\Phi,\{\mu\},\l)\cdot 
\left(g^{(1)}\right)^{-1}(F_1,\mu_1,\l)
e^{-\P_1} \cdot\, \ldots \, \cdot
\left(g^{(1)}\right)^{-1}(F_l,\mu_l,\l)
e^{-\P_l}\0\\
l&=&0,1,\ldots ,N \0\\
g^{(N)}_0(\Phi,\{\mu\},\l)&=&g^{(N)}
(\Phi,\{\mu\},\l)\label{4.7}
\b
Taking into account (\ref{4.1a}), 
(\ref{4.1b}) and (\ref{4.5}) we observe
 that 
the above element can be alternatively 
expressed as
\a
&&g^{(N)}_l(\Phi,\{\mu\},\l)=
 \sum_{r\in {\ZZ}_{n+1}}
S^r |v_{\Phi}( \{\mu\},\o^{-r}\l)>
<\rho_l(\o^{-r}\l)|
 S^{-r}\cdot e^{-K(F_l)-\P_l}\0\\
&&<\rho_l(\l)|=\sum_{j=1}^{n+1}
<j|\rho_{jl}(\l) \label{4.8}
\b
where the coefficients $\rho_{jl}$ 
satisfy the following recursion 
relations
\a
&&\rho_{jl+1}(\l)=\frac{1}{n+1}
\sum_{k=1}^{n+1}e^{L_{kl+1}}
\rho_{kl}(\l)
\sum_{s\in{\ZZ}_{n+1}} \o^{(j-k)s} 
\frac{\l-\o^s\mu_{l+1}e^{f_{kl+1}}}
{\l-\o^s\mu_{l+1}}\0\\
&&L_{kl+1}=K_k(F_{l+1})-K_k(F_l)-
\frac{f_{kl+1}+p_{kl}}{2}\label{4.9}
\b
together with the initial
 conditions (see (\ref{4.1a}), (\ref{4.1b}))
\a
\rho_{j0}&=& \frac{1}{n+1} \label{4.10}
\b

To fix recursively the unknown abelian
 factors $F_l$ and $\P_l$ we shall
 impose the following conditions:
 first, the element $g^{(N)}_l$ (\ref{4.8})
 has  no poles at the points 
$\l=\mu_1,\, \ldots ,\, \mu_l$
\a
{\rm res}_{\mu_i} g^{(N)}_l&=&0
,~~~~~~1\leq i \leq l \label{4.11}
\b
and second, we require that the
 multipliers $\rho_{jl}$ (\ref{4.8}),
 (\ref{4.9}) are meromorphic functions 
on $\l$ which have  simple poles at 
$\l=\o^{r_i}\mu_i$ for $i=1,\, \ldots ,\, l$ 
and are homolomorphic elsewhere. Due to
 this restriction we can write
\a
{\rm res}_{\o^r \mu_i}\rho_{jl}
&=&0,~~~~~r\neq r_i,\,\,\, 1\leq i \leq l
\label{4.12}
\b
The discrete parameters 
$r_i=1,\, \ldots ,\, n$ appeared naturally
 in the description of the soliton solutions
 (\ref{2.33}), (\ref{2.34}), (\ref{3.7}).
We stress that due to (\ref{3.16}), the
 requirement (\ref{4.11}) guarantees
 that the matrix $g^{(N)}_l$ (\ref{4.7})
 has  no singularities for $\l^{n+1}=\mu_i^{n+1}$,
 $i=1,\, \ldots , \, l$. The significance
 of (\ref{4.12})
will become clear in what follows.
 Taking into account (\ref{4.12}), we see that
 (\ref{4.11}), written in terms of the
 entries of the matrix (\ref{4.7}),
 (\ref{4.8}), reads
\a
{\rm res}_{\mu_l} \left( g^{(N)}_{l+1}\right)_{ij}
&=&0\0\\
&\Updownarrow&\0\\
 {\rm res}_{\o^{r_{l+1}}
\mu_{l+1}} \rho_{jl+1}&=&
\mu_{l+1}(1-\o^{r_{l+1}})
\o^{(j+1-i)r_{l+1}}\times\0\\
\times \prod_{a\neq l+1} 
\frac{\o^{r_{l+1}}\mu_{l+1}-\mu_a}
{\mu_{l+1}-\mu_a}
&\cdot &
\prod_a \frac{ \mu_{l+1}+\ep_{ia}}
{\o^{r_{l+1}}\mu_{l+1}+\ep_{ia}}
\rho_{jl+1} (\mu_{l+1})
\label{4.13}
\b
The last equation is consistent since due
 to (\ref{3.7}), its r. h. s. does not depend 
on $i$. Therefore, we conclude that the 
soliton dynamics, encoded in (\ref{3.7})
 is crucial in solving the factorization
 problem (\ref{4.2}). This
observation also explains the reason 
to impose the condition (\ref{4.12}). 
On the other hand, in view of 
(\ref{4.9}) we get
\a
{\rm res}_{\o^r\mu_{l+1}}\rho_{jl+1}&=&
\frac{\o^{jr} \mu_{l+1}}{n+1}
\sum_{k=1}^{n+1}\o^{r(1-k)}
e^{L_{kl+1}}\rho_{kl}
(\o^r\mu_{l+1}) (1-e^{f_{kl+1}})\0\\
r&\in & {\ZZ}_{n+1} \label{4.14}
\b
where the factors $L_{kl+1}$ were 
defined by (\ref{4.9}). Inserting 
the above expression into (\ref{4.13})
 we get another consistency condition
\a
\rho_{jl+1}(\mu_{l+1})&=&
\rho_{kl+1}(\mu_{l+1})\0\\
j,k&=&1,\, \ldots , \, n+1\label{4.15}
\b 
To calculate the above quantities we
 recall the general identity
\ai
\frac{1}{n+1}\sum_{r\in {\ZZ}_{n+1}} 
\frac{\o^{rk}}{\l-\o^{-r}\mu}&=&
\frac{\l^{n-k}\mu^k}{\l^{n+1}-\mu^{n+1}}\0\\
0\leq & k &\leq n 
\label{4.16a}
\b
from which in the limit 
$\frac{\l}{\mu} \rightarrow 1$ one gets
\a
\sum_{r=1}^{n}
 \frac{\o^{rk}}{1-\o^r}&=&\left\{ \begin{array}{cc}
\frac{n}{2}+k& -n\leq k \leq 0\\

\\

-\frac{n}{2}-1+k& 1\leq k \leq n+1 
\end{array}\right. \label{4.16b}
\bj
Setting $\l=\mu_{l+1}$ in (\ref{4.9}) 
and using (\ref{4.12}) for $r=0$, we
 obtain with the help of (\ref{4.16b})
 the following expressions
\a
&&\rho_{jl+1}(\mu_{l+1})=e^{L_{jl+1}}
 \rho_{jl}(\mu_{l+1})+\frac{\mu_{l+1}}{n+1}
\sum_{k=1}^{n+1} e^{L_{kl+1}}
(1-e^{f_{kl+1}})\rho^{'}_{kl}(\mu_{l+1})-\0\\
&&-\frac{1}{n+1}\left( \sum_{k=1}^{n+1}(k-j) 
+(n+1)\sum_{k=1}^j\right) 
e^{L_{kl+1}}\rho_{kl}(\mu_{l+1})
(1-e^{f_{kl+1}})\label{4.17}
\b
Inserting this identity in (\ref{4.15})
 and taking into account the second equation
 (\ref{4.9}) we obtain
\a
e^{\frac{p_{jl}}{2}}&=&\frac{ \rho_{jl}
(\mu_{l+1})}{\left(\prod_k \rho_{kl}
(\mu_{l+1})\right)^{\frac{1}{n+1}}}
 e^{-K_j(F_l)}\label{4.18}
\b
The j--independent factor in the 
denominator is fixed by (\ref{4.3}). 
Therefore, the algebraic system (\ref{4.12}), 
(\ref{4.13})  reduces to 
\a
&&\sum_{k\in {\ZZ}_{n+1}} \o^{r(1-k)}
\left( \frac{\rho_{kl}(\o^r\mu_{l+1})}
{\rho_{kl}(\mu_{l+1})}-\o^{-r} \frac{\rho_{k+1l}
(\o^r\mu_{l+1})}
{\rho_{k+1l}(\mu_{l+1})}\right)e^{K_k(F_{l+1})-
\frac{f_{kl+1}}{2}}=\delta_{r, r_{l+1}}
(1-\o^r)\times\0\\
&&\times \prod_{a\neq l+1} 
\frac{\o^r\mu_{l+1}-\mu_a}
{\mu_{l+1}-\mu_a}
\prod_a \frac{ \mu_{l+1}+\ep_{1a}}
{\o^r\mu_{l+1}+\ep_{1a}}
\times\0\\
&& \times\sum_{k\in {\ZZ}_{n+1}}
\left( 1+\mu_{l+1} \frac{d}{d \l}
\ln \frac{\rho_{kl}}{\rho_{k+1l}}
(\mu_{l+1})\right)e^{K_k(F_{l+1})-
\frac{f_{kl+1}}{2}}\label{4.19}
\b
Note that for $r=0$ the above equation
 is satisfied identically. We recall also
 that $r_j\neq 0 \,  {\rm mod}\, (n+1)$.
 Due to this and since $K(F_{l+1})$ and
$F_{l+1}$ are traceless, we conclude
 that (\ref{4.19}) determines uniquely 
$F_{l+1}$ as a function of $F_l$. 
Continuing this procedure, we
 finally 
 arrive at the element $g_N^{(N)}$ 
(\ref{4.7}). From (\ref{4.11}) we
 conclude that it is 
a {\it holomorphic} function on 
 the spectral parameter. 
 This wants to say that $g_N^{(N)}$
 does not depend on $\l$. Due
 to the fact that this element
 satisfies (\ref{3.16}), it is 
clear that its off--diagonal
 elements
vanish identically. The unique 
element which remains undetermined 
is $\P_N$
(\ref{4.3}). One can use this
 ambiguity to set $g_N^{(N)}=1$.
 This completes
the factorization procedure.

In what follows we shall
 concentrate our attention
 on the \dg\, elements which
 generate monosolitons from 
the vacuum. Due to the general
 expressions and in view of the
 of the one--soliton specification
 (\ref{4.4}), one gets
\ai
\hskip -1.5cm \frac{\del}{\del \var_i}
g^{(1)}(\Phi , \mu , \l)\cdot
\left(g^{(1)}\right)^{-1}
(\Phi,\mu,\l)&=&e^{K(\Phi)}
 \left( B^i(\mu,\l)-B^{n+1}(\mu,\l)\right)
e^{-K(\Phi)}\label{4.20a}\\
&i=&1,\, \ldots , \, n\0\
\b
where $\Phi$ (\ref{2.23}) is an
 one--soliton solution (\ref{2.43}), $K(\Phi)$
was introduced in (\ref{4.5}), (\ref{4.6}) and 
\a
B^i(\mu , \l)&=&\sum_{l<i} \frac{\l^{n+1+l-i}
\mu^{i-l}}{\l^{n+1}-\mu^{n+1}}E^{il}+
\frac{1}{2}
\frac{\l^{n+1}+\mu^{n+1}}
{\l^{n+1}-\mu^{n+1}}E^{ii}+\0\\
&+&\sum_{l>i} \frac{\l^{l-i}
\mu^{n+1+i-l}}{\l^{n+1}-
\mu^{n+1}}E^{il},~~~~
i=1,\, \dots ,\, n+1
\label{4.20b}
\bj
We recall that since $\Phi$ is 
traceless, one of its components can be 
expressed in terms of the others. In 
calculating the derivatives in the l. h. s. 
of (\ref{4.20a}) we have set 
$\var_{n+1}=-\sum_{i=1}^n\var_i$. The
 expressions (\ref{4.20b}) follow from 
(\ref{4.1a}), (\ref{4.1b}), (\ref{4.4}) 
and the summation formula (\ref{4.16a}).
Let us introduce the loop group element
\a
h^{(1)}(\Phi , \mu , \l)&=&e^{-K(\Phi)} 
g^{(1)}(\Phi , \mu , \l)\label{4.21}
\b
From (\ref{4.20a}) it follows that this 
element satisfies the system
\ai
\frac{\del}{\del \var_i}h^{(1)}(\Phi , 
\mu , \l)&=& J^i(\mu,\l)  
h^{(1)}(\Phi , \mu , \l)\0\\
i&=&1,\, \ldots ,\, n\label{4.22a}\\
J^i(\mu , \l)&=&\left(\frac{\del}
{\del \var_{n+1}}- \frac{\del}
{\del \var_{i}}\right)K(\Phi)+
B^i(\mu,\l)-B^{n+1}(\mu ,\l)\0\\
\var_{n+1}&=& -\sum_{i=1}^n \var_i 
\label{4.22b}
\bj
An intriguing property of the loop 
algebra elements $J^i$ is that they 
{\it do not depend} on the affine Toda
 fields $\var_i$ for $i=1,\ldots , n$.
Combining this observation with the
 integrability condition of the 
linear differential system (\ref{4.22a})
 we conclude that 
\a
\left[ J^i(\mu , \l), J^j (\mu , \l)\right]&=&0 
\label{4.23}
\b
and therefore the following representation
\a
g^{(1)}(\Phi , \mu , \l)&=&e^{K(\Phi)}
e^{\sum_{i=1}^n \var_i J^i(\mu ,\l)}
\label{4.24}
\b
takes place. Note that for $n=1$, which 
corresponds to the \shG\, model, the 
diagonal prefactor in the r. h. s. of the
 above equation disappears and we 
end up with the exponentiated form of 
the one--soliton \dg\, element \cite{BB}.

We proceed with the following remark: 
as it was noted in \cite{fr}, the general
 dressing problem admits two solutions
 depending on the analiticity properties 
for $\l \rightarrow 0$ and 
$\l \rightarrow \infty$. In other words, 
the solutions 
(\ref{3.24a}), (\ref{3.24b}) in particular,
 due to the Gauss decomposition 
in the loop group $\Sll$, represent two
different elements. Skipping the dependence
 on the field parameters one can write
\a
g(\l)&=&
\left\{ \begin{array}{cc} g_+(\l)& 
\l\rightarrow 0\\

\\

g_-(\l)& \l\rightarrow \infty 
\end{array}\right. \label{4.25}
\b
from where we conclude that the 
element $g(\l)$  we analyzed before
is an analytic continuation of two
 {\it different} elements of the loop group.
As noted in \cite{fr}, \cite{STS}, a \dg\,
 element is represented by the pair 
$(g_+, g_-)$ and there is a canonical 
diffeomorphism between the dressing group
 and the underlying loop group
\footnote{We stress that this map is not an
 isomorphism of Lie groups since it does not 
preserve the multiplication}
\a
(g_+,g_-) \rightarrow g_-^{-1}g_+ \label{4.26}
\b
Denote by $J^i_+(\mu,\l)$ and $J^i_-(\mu,\l)$ the
expansions of the elements (\ref{4.22b}) around
$\l \rightarrow 0$ and $\l \rightarrow \infty$ 
respectively.
Let us calculate the value of the map (\ref{4.26})
 for the one--soliton \dg\, elements.
Expanding (\ref{4.20b}) around
 $\l \rightarrow  0$ and $\l \rightarrow \infty $
 and  using (\ref{4.22b}) we get (\ref{4.24})
\a
&&(g_-^{(1)})^{-1}(\Phi ,\mu , \l)
g_+^{(1)}(\Phi , \mu , \l)=
:e^{\sum_{i=1}^n \var_i I^i(\mu)}:\0\\
&& \0\\ 
&&I^i(\mu)=J^i_+(\mu,\l)-J^i_-(\mu,\l)=\0\\ 
&&=-\sum_{l=1}^{n+1} 
\sum_{k\in {\ZZ}}
\left( \frac{E^{il}_{l-i+k(n+1)}}
{\mu^{k(n+1)+l-i}}-
 \frac{E^{n+1l}_{l+k(n+1)}}
{\mu^{k(n+1)+l}}\right),~~~i\in {\ZZ}_{n+1}
\label{4.27}
\b
where  the normal product
 $: \,\,\, :$ means writing $J^i_-$ 
on the left $: J^i_-J^j_+:=:J^j_+J^i_-:=J^i_-J^j_+$ and
the lower indices, as we defined in
 Sec. 2, count the powers of $\l$. 
To be more precise, one should stress 
that the elements $I^i$ are not well defined 
in the loop algebra. To avoid this difficulty,
one considers the level one representations 
of the corresponding affine Lie algebra \cite{Olive}, 
\cite{Cat}, \cite{Kac}, \cite{al}. 
It is not difficult to calculate the
 commutators of the above elements with
the grade $\pm 1$ elements of the principal
 Heisenberg subalgebra (\ref{Heis}).
The result is
\a
\left[ I^i(\mu) , \E_+ \right]&=&
\mu\left( I^i(\mu) -I^{i-1}(\mu)+
 I^n(\mu)\right)\0\\
\left[ I^i(\mu) , \E_- \right]&=& 
\frac{1}{\mu}
\left( I^i(\mu)-I^{i+1}(\mu)+
I^1(\mu)\right)\label{4.28}
\b
Comparing the expressions (\ref{4.27}) 
with (\ref{2.16a}), (\ref{2.16b}) and 
(\ref{2.21}) we obtain
\a
I^i(\mu)&=& \sum_{l\in {\ZZ}_{n+1}} \o^l
 (\o^{-il}-1)F^l(\mu)
\label{4.29}
\b
The above identity together with (\ref{2.22}),
 (\ref{4.27}) and (\ref{4.28})
suggests a relation to the vertex operator 
approach to the affine Toda solitons \cite{Olive}, 
\cite{Cat}.

It is worthwhile to make several remarks 
and comments. First of all, we recall that there 
are two related but inequivalent notions of
solitons \cite{Raj}--\cite{Fad}. From the physical
 point of view a soliton is a localized solution
 of the field equations which carries finite 
physical quantities, like momentum, energy, etc.
 There is another concept of solitons, adopted
 within the ISM, namely solitons arise when 
the underlying auxiliary linear problem is 
reflectionless. In the present paper we
 relax the physical requirements and deal 
with the solitons 
 as they were treated by the ISM. Our reason
 to do this is that the formalism developed by us 
can be repeated without any modification in
 the physical region of the coupling constant.
 Due to this, for the sake of brevity, we
 preferred to work with certain real value 
of the coupling constant and to enjoy the 
algebraic beauty of the 
soliton solutions. Second, in the present
 paper we restricted ourselves to study $\An$
 Toda solitons only. One reason to do that was
 our intention to keep the discussion as
 elementary as it is possible. On the other
 hand, our approach is based on the
 observation that the connection (\ref{2.27})
 belongs to the principal gradation. Due to
 that we exploited the observation that the
 gradation generating automorphism $\s$ 
(\ref{2.5}) yields a symmetry of the linear
 system (\ref{2.28}). For general Lie algebras,
 one can repeat the procedure 
of Sec. 2 to construct soliton
 solutions, but in general it is
 not possible to construct the
 counterpart of (\ref{2.31}). The reason 
is that the order of $\s$
 (\ref{2.5}), except for the $A_n$ and $C_n$ 
Lie algebras, is always smaller than
 the dimension of any irreducible
 representation. Therefore,
  to generalize the results of
 Sec. 3 and Sec. 4, we have to look
 for an additional symmetry of the 
linear system (\ref{2.28}). This problem 
is under investigation. Finally, we note
 that in contrast to the seminal paper
 \cite{BB}, where in the particular
 example of the \shG\, model the
 factorization problem (\ref{4.7})
 was treated by using B\"{a}cklund
 transformations, we preferred the
 algebraic recursive approach
 described in the present Section.
 It remains as an open question to
 relate these two approaches. 
The B\"{a}cklund transformations
 for $\An$ Toda equations have been
 studied in \cite{LOT}. As final 
comment, we stress that the expressions
(\ref{4.27})--(\ref{4.29}) provide the relation 
to the vertex operator formalism only for 
one soliton solutions. We hope to go back
to this problem for general N--solitons 
elsewhere.

\vskip 2cm

{\bf Acknowledgements} It is a
 pleasure to thank 
L. A. Ferreira,
 M. A. C. Kneipp and J. P. Zubelli
 for stimulating discussions and various 
comments on the preliminary version  of this paper.
 R. P. is grateful
 to L. Bonora for bringing his attention on the 
ref's \cite{Date} and \cite{Nid}. The authors would like to 
thank J. A. Helay\"el--Neto for the careful 
reading of the manuscript. We gratefully
 acknowledge the financial support from CNPq--Brazil.


\begin{thebibliography}{}
\bibitem{Raj} R. Rajaraman, {\it Solitons and
 Instantons}, North--Holland, Elsevier 1982.
\bibitem{FaKo} L. D. Faddeev and
 V. E. Korepin, Phys Rep. ${\bf 420}$(1978)1--78.
\bibitem{Ab} M. J. Ablowitz and H. Segur, 
{\it Solitons and the Inverse Scattering Method}, 
Siam, Philadelphia, 1981.
\bibitem{Nov}S. Novikov, S. Manakov, 
L. P. Pitaevsky and V. E. Zakharov, 
{\it  Theory of Solitons}, Consultants Bureau, 1984.
\bibitem{Fad} L. D. Faddeev and 
L. A. Takhtajan, {\it Hamiltonian Methods 
in the Theory of Solitons}, Springer, 1987.
\bibitem{T} M. Toda, Phys. Rep. ${\bf 18}$(1978)1.
\bibitem{OT} D. Olive and N. Turok,
 Nucl.${\bf B215}$(1983)470; 
Nucl. Phys. ${\bf B220}$(1983)491-507.
\bibitem{DS} V. G. Drinfeld and V. Sokolov, 
Sov. Math. Dokl. ${\bf 23}$(1981)457.
\bibitem{gDS} N. Burroghs, M. de Groot, 
T. Hollowood and L. Miramontes,
Phys. Lett. ${\bf B277}$(1992)89--94\\
 M. de Groot, T. Hollowood and L. Miramontes, 
Commun. Math. Phys. ${\bf 145}$(1992)57--84\\
 N. Burroghs, M. de Groot, T. Hollowood and 
L. Miramontes,, Commun. Math. Phys.
 ${\bf 153}$(1993)187--215.
\bibitem{eng}D. Olive and N. Turok, 
Nucl. Phys. ${\bf B257}$(1985)277--301;  
Nucl.Phys. ${\bf B265}$(1986)469--484.
\bibitem{ZS} V. Zakharov and A. Shabat, 
Funct. Anal. Appl. ${\bf 13}$(1979)166.
\bibitem{jap} E. Date, M. Jimbo, M. Kashiwara 
and T. Miwa, {\it Transformation Groups for
 Soliton Equations}, in: Non--linear integrable
 systems, eds M. Jimbo and T. Miwa 
( World Scientific, Singapore, 1983).
\bibitem{fr}O. Babelon, D. Bernard, Phys.
 Lett. ${\bf B260}$ (1991) 81,\\
Commun. Math. Phys. ${\bf 149}$ (1992) 279.
\bibitem{STS} M. Semenov--Tian--Shansky,
 Publ. RIMS ${\bf 21}$ (1985)
1237,\\
M. Semenov--Tian--Shansky,
 {\it Poisson Lie Groups, Quantum Duality
Principle, and the Quantum Double},
 hep--th/9304042.
\bibitem{GR} G. Cuba and R. Paunov, Phys.
 Lett. ${\bf B381}$(1996)255--261.
\bibitem{ago} L. A. Ferreira, L. Miramontes 
and J. S. Guill\'en,
{\it Tau--functions and Dressing Transformations
 for Zero--Curvature Affine Integrable Equations},
 hep--th/9606066 and J. Math. Phys., to appear.
\bibitem{Hol}T. Hollowood, Nucl. 
Phys. ${\bf B384}$(1992)523-548.
\bibitem{Clis} H. Aratyn, C. P. Constantinidis,
 L. A. Ferreira, J. F. Gomes and A. H. Zimerman, 
Nucl. Phys. ${\bf B406}$(1993)727\\
N. J. MacKay and W. A. McGhee, Int. J. Mod. Phys. 
${\bf A8}$(1993)2791--2807\\
Z. Zhu and D. G. Caldi, Nucl. Phys. 
${\bf B436}$(1995)659--678.
\bibitem{Olive} D. I. Olive, M. W. Saveliev and
 J. W. Underwood, Phys. Lett 
${\bf B311}$(1993)117--122\\
D. I. Olive, N. Turok and J. W. Underwood,
 Nucl. Phys. ${\bf B401}$(1993)663--697;
Nucl. Phys. ${\bf B409}$(1993)509--546\\
M. A. C. Kneipp and D. I. Olive, Nucl. Phys. 
${\bf B408}$ (1993) 565--578\\
M. A. C. Kneipp, Ph. D. thesis, 
Swansea, 1995.
\bibitem{Cat}M. A. C. Kneipp and
 D. I. Olive, Commun. Math. Phys.
 ${\bf 177}$ (1996)561--582.
\bibitem{body} O. Babelon and D. Bernard, 
Phys Lett. ${\bf B317}$(1993)363\\
H. Braden and A. N. W. Hone, Phys. Lett. 
${\bf B380}$(1996)296--302.
\bibitem{dual} E. Martinec and N. Warner,
 Nucl. Phys. ${\bf B459}$(1996)97--112\\
R. Donagi and E. Witten, Nucl. 
Phys. ${\bf B460}$(1996)299-334\\
I. M. Krichever and D. H. Phong, 
{\it On the Integrable Geometry of Soliton
 Equations and $N=2$  Supersymmetric Gauge Theories},
 hep-th/9604199\\
P. M. Sutclife, Phys. Lett ${\bf B381}$(1996)130.
\bibitem{Date}E. Date, Osaka J. Math, 
${\bf 19}$(1982)125.
\bibitem{Nid} M. Niedermaier,
 Commun. Math. Phys. ${\bf 160}$ 391--429 (1994).
\bibitem{BB} O. Babelon and D. Bernard,
 Int. J. Mod. Phys. ${\bf A8}$ (1993)507.
\bibitem{Kac}V. G. Kac, {\it Infinite 
dimensional Lie algebras}.Third edition.
Cambridge Univ. Press. 1990.
\bibitem{al} G. G. A. B\"auerle and E. A. de Kerf, 
{\it Lie Algebras}, Part 1,
 Stud. in Math. Phys. ${\bf 1}$, North--Holland,
 Amsterdam.
 \bibitem{BJ} E. J. Beggs and P. R. Johnson,
 {\it Inverse Scattering and Solitons in
 $A_{n-1}$ Affine Toda Field Theories}, 
 hep--th/9610104 and Nucl. Phys. ${\bf B}$,
 to appear
\bibitem{FMc} H Flaschka and D. W. McLaughlin, 
Prog. Theor. Phys, ${\bf 55}$(1976)438.
\bibitem{R}R. Paunov, Phys. Lett 
${\bf B347}$(1995)63.
\bibitem{LS} A. N. Leznov and M. V. Saveliev, 
Progress in Physics ${\bf 15}$, Birkhauser, 
Basel, 1992.
\bibitem{cat} O. Babelon and L. Bonora,
 Phys. Lett. ${\bf B244}$(1990)220.
\bibitem{SP}H. Aratyn, L. A. Ferreira, J. F. Gomes and
A. H. Zimerman, Phys. Lett. ${\bf B254}$(1991)372--380\\
C. P. Constantinidis, L. A. Ferreira, J. F. Gomes and 
A. H. Zimerman, Phys. Lett. ${\bf B298}$(1993)88--94.
\bibitem{LOT}H. C. Liao, D. Olive and 
N. Turok, Phys. Lett. ${\bf B298}$(1993)95--102.
\end{thebibliography}
\end{document}